\documentclass[aps,prd,preprint,groupedaddress,showpacs,floatfix]{revtex4}%
\usepackage{graphicx}
\usepackage{amsmath}
\usepackage{bm}

\begin{document}

\preprint{ANL-HEP-PR-02-021}

\title{Hadronic decays of $\bm{\chi_{bJ}}$ into light bottom squarks} 


\author{Edmond L. Berger and Jungil Lee}
\affiliation{
High Energy Physics Division, 
Argonne National Laboratory, Argonne, IL 60439
}


\date{\today}

\begin{abstract}
We calculate the rates for inclusive hadronic decay of the three $\chi_{bJ}$ 
states into a pair of light bottom squarks as a function of the masses of the 
bottom squark and the gluino.  We include color-singlet and color-octet 
configurations.  The color-octet contribution is found to be insignificant for 
the $\chi_{b0}$ but can dominate in the $\chi_{b2}$ case if current lattice estimates 
are used for the color-octet matrix element.  In comparison with the standard 
model values, bottom squark decays can increase the predicted hadronic width of 
the $\chi_{b0}$ by as much as 33\%, for very small bottom squark masses and gluino 
masses in the range of 12 GeV, but make a small contribution in the cases of 
$\chi_{b1}$ and $\chi_{b2}$.  Data from decays of the $\chi_{bJ}$ states could 
provide significant new bounds on the existence and masses of supersymmetric 
particles.  
\end{abstract}

\pacs{12.38.Bx, 13.85.Ni, 13.87.Fh, 14.40.Gx}

\maketitle

\section{Introduction\label{intro}}
The possible existence of a light bottom squark $\tilde{b}$ with mass similar 
to or less than that of the bottom quark $b$ is investigated in several theoretical 
papers~\cite{BHKSTW,CHWW,DedesDreiner,Kagan,loops1,Nierste,elblc}.  This 
hypothesis is not inconsistent with direct experimental searches and 
indirect constraints from other observables~\cite{CLEO,DELPHI,ELBrev}.  If 
the mass $m_{\tilde{b}}$ is 
less than half of that of the $P$-wave $\chi_b$ resonances, then 
the direct decay $\chi_b \rightarrow \tilde{b} \tilde{b}^*$ could 
proceed with sufficient rate for observation, particularly in data 
being accumulated at the Cornell CESR facility~\cite{CLEO-c}.  In this paper, 
we compute the expected rates for decays of the three $\chi_{bJ}$ states into a 
pair of bottom squarks as a function of the masses of the bottom 
squark and the gluino.  The mass of the gluino $\widetilde{g}$ 
enters because the gluino is exchanged in the relevant decay subprocesses. 
The $\tilde g$ and the $\tilde b$ are the spin-1/2 and spin-0 supersymmetric 
(SUSY) partners of the gluon ($g$) and bottom quark.  This paper extends previous  
work of one of us~\cite{elblc} on decays of the Upsilon states,
$\Upsilon(nS) \rightarrow \tilde{b} \tilde{b}^*, n = 1 - 4$.

In the remainder of this Introduction, we summarize the phenomenological 
motivation and status of the hypothesis of a light bottom squark 
along with a light gluino and then present an outline of the remainder of the 
paper.  A recent presentation of the experimental properties of the 
$\chi_{bJ}$ states may be found in Ref.~\cite{CLEO-chi}.  
 
The hypothesis of a relatively light color-octet gluino $\tilde g$ (mass 
$\simeq 12$ to 16 GeV) that decays with 100\% branching fraction into a bottom 
quark $b$ and a light color-triplet bottom squark $\tilde b$ (mass $\simeq 2$ to 
5.5 GeV) is proposed in Ref.~\cite{BHKSTW} in order to explain the observed excess 
cross section for production of bottom quarks at hadron colliders.   In this 
scenario $\tilde b$ is the lightest SUSY particle(LSP), and the masses of all other 
SUSY particles are arbitrarily heavy, i.e., of order the electroweak scale or 
greater.  Improved agreement is obtained with the magnitude and shape of the transverse 
momentum distribution of bottom-quark production at hadron colliders.  The proposal is 
consistent with measurements of the time-averaged $B^0 \bar{B}^0$ mixing 
probability~\cite{BHKSTW}.  The $\tilde b$ either picks up a light quark, becoming a 
spin-1/2 $``$mesino", perhaps living long enough to escape from a typical collider detector, 
or it decays promptly via baryon-number R-parity violation into a pair of hadronic 
jets~\cite{BHS,ELBrev}.  

There are important restrictions on the existence and couplings of bottom squarks 
from precise measurements of $Z^0$ decays.  A light $\tilde b$ would be ruled out 
unless its 
coupling to the $Z^0$ is very small.  The squark couplings to the $Z^0$ depend 
on the mixing angle $\theta_{\tilde{b}}$.  If the light 
bottom squark ($\widetilde{b}_1$) is an appropriate mixture of left-handed and 
right-handed bottom squarks, its lowest-order (tree-level) coupling to the $Z^0$ 
can be arranged to be small~\cite{CHWW} if $\sin^2 \theta_{\tilde{b}} \sim 1/6$.  
The exclusion by the CLEO collaboration~\cite{CLEO} of a $\tilde b$ with mass 
3.5 to 4.5 GeV does not apply since their analysis focuses only on the 
leptonic decays $\tilde b \rightarrow c \ell \tilde \nu$ and 
$\tilde b \rightarrow c \ell $.  The $\tilde b$ need not decay leptonically nor 
into charm.  The search by the DELPHI collaboration~\cite{DELPHI} for long-lived 
squarks in their $\gamma \gamma$ event sample is not sensitive to 
$m_{\tilde b} < 15$ GeV.  Bottom squarks make a small contribution to the
ratio $R$ of the inclusive cross section for $e^+ e^- \rightarrow$ hadrons 
divided by that for $e^+ e^- \rightarrow \mu^+ \mu^-$, requiring an 
accuracy for detection at the level of 2\% or so, much greater than that of   
current measurements in the 6 to 7\% range~\cite{BES}.  In $e^+ e^-$ production, 
resonances in the $\tilde{b} {\tilde{b}}^*$ system are likely to be impossible 
to extract from backgrounds~\cite{Nappi}, but $\gamma \gamma$ collisions may 
be more promising.  The angular distribution of hadronic jets produced in 
$e^+ e^-$ annihilation can be examined in order to bound the contribution of
scalar-quark production.  Spin-1/2 quarks and spin-0 squarks emerge with
different distributions, $(1 \pm {\rm cos}^2 \theta)$, respectively. 
The angular distribution measured by the CELLO
collaboration~\cite{CELLO} is consistent with the production
of a single pair of charge-1/3 squarks along with five flavors of
quark-antiquark pairs.  

The possibility that the gluino may be much less massive than most other 
supersymmetric particles is intriguing from different points of 
view~\cite{Fayet,Farrar,LC2,Raby}. 
An early study by the UA1 Collaboration~\cite{ua1gluino} 
excludes $\tilde{g}$'s in the mass range $4 < m_{\tilde{g}} 
< 53$ GeV, but it starts from the assumption that there is a light neutralino 
${\tilde{\chi}}_1^0$ whose mass is less than the mass of the gluino.  The 
conclusion is based on the absence of the expected decay $\tilde{g} \rightarrow 
q+\bar{q}+{\not \! E_T}$, where $\not \! E_T$ represents the missing energy 
associated with the ${\tilde{\chi}}_1^0$.  In the scenario discussed here, this 
decay process does not occur since the bottom squark is the LSP and the 
${\tilde{\chi}}_1^0$ mass is presumed to be large 
({\em i.e.}, $> 50$ GeV). An analysis of 2- and 4-jet events by the ALEPH 
collaboration~\cite{ALEPH} disfavors $\tilde g$'s with mass 
$m_{\tilde g} < 6.3$ GeV but not $\tilde g$'s in the mass range relevant for the 
SUSY interpretation of the bottom quark production cross section.  A similar 
analysis is reported by the OPAL collaboration~\cite{OPALg}.  A light 
$\tilde{b}$ is not excluded by the ALEPH analysis.  A very precise measurement 
of the $\beta$ function in QCD is potentially the most sensitive probe of the 
existence of a light gluino since the color-octet gluino makes a contribution to 
$\beta$ equivalent to that of 3 new flavors of quarks.  However, the analysis of 
data must take into account the effects of SUSY-QCD production processes.  The 
combined ranges of $\tilde{b}$ 
and $\tilde{g}$ masses proposed in Ref.~\cite{BHKSTW} are compatible 
with renormalization group equation constraints and the absence of color 
and charge breaking minima in the scalar potential~\cite{DedesDreiner}.  

In Sec. II, we begin with a discussion of factorization.  The $\chi_b$ is treated 
as a $b \bar{b}$ bound system and described non-perturbatively.  The transition 
of the (on-shell) $b \bar {b}$ system into the observed final state is 
calculated in QCD perturbation theory.  In Sec. II, we also outline the 
projection of the on-shell $b \bar{b}$ system into the states of spin, parity, 
and angular momentum of interest.  We present our explicit calculation of 
$\chi_b \rightarrow \tilde{b} \tilde{b}^*$ in Sec. III.  We include both 
color-singlet and color-octet configurations.  Results and conclusions are 
summarized in Sec. IV.  When compared with the leading-order expectation in 
the standard model, the bottom squark contribution can increase the 
predicted hadronic width of the $\chi_{b0}$ by as much as 33\% for small 
bottom squark masses, $O(2$ to 4 GeV), and gluino masses in the range of 12 GeV.  
However, bottom squark decays make a negligible contribution in comparison with 
the standard model value in the $\chi_{b1}$ and $\chi_{b2}$ cases.  The color-octet 
contribution to bottom squark decays is insignificant for the $\chi_{b0}$ but can 
dominate in the $\chi_{b2}$ case if current lattice estimates are used for the 
color-octet matrix element.  Data from decays of the $\chi_b$ states could provide 
significant new bounds on the existence and masses of supersymmetric particles.  
 

\section{Factorization and Conventional Decays of the $\chi_{bJ}$}
To compute the decay of a heavy quarkonium state, we begin with the 
statement of factorization.  The initial $\chi_b$ is taken to be 
a bound state $b \bar{b}$ of bottom quarks.  To the extent that the 
$b$ quarks are heavy enough, the bound state aspects may be described 
with non-relativistic quantum chromodynamics(NRQCD)~\cite{CasLeP}.  The 
decay dynamics are assumed to be described by short-distance perturbative 
QCD, in turn also justified in part by the large mass of the $b$ 
quark.
The decay rate of the heavy quarkonium state is then expressed in 
the factored form~\cite{BBL-P,BBL}
\begin{eqnarray}
\Gamma(\chi_{bJ})
=\sum_n C_n\langle \chi_{bJ}|{\cal O}_n|\chi_{bJ}\rangle.  
\label{eq:factorization}
\end{eqnarray}
The summation index $n$ stands for the spectroscopic state 
$^{2S+1}L_J^{(1,8)}$ of the $b\overline{b}$ pair. The superscript 
to the right $(1,8)$ stands for the color quantum number, 
singlet~\cite{P-LO,Berger:1980ni} and octet~\cite{BBL-P}, respectively.  
The short-distance coefficient $C_n$, describes the inclusive transition 
of an on-mass-shell $b \bar{b}$ system into the relevant final state, 
and it is calculable perturbatively in a series in the strong coupling
strength $\alpha_s(m_b)$.  The non-perturbative matrix element 
$\langle\chi_{bJ}|{\cal O}_n|\chi_{bJ}\rangle$
represents the probability that a $\chi_{bJ}$ meson
evolves into a free $b\overline{b}$ pair with quantum numbers $n$. 
The four quark operator $\mathcal{O}_n$ is defined in Ref.~\cite{BBL}.

Although the recent CDF measurement of the polarization of prompt 
$J/\psi$'s presents a serious challenge for the NRQCD factorization 
formalism for inclusive charmonium production~\cite{PSI-POL}, factorization 
in the bottomonium case is believed to be safe since $m_b$ is well 
separated from the long distance scale $m_b v^2_b$.  For example, the 
recent CDF measurement of $\Upsilon(nS)$ polarization agrees with the 
NRQCD prediction~\cite{UPS-POL} based on matrix elements fitted to the 
Run IB data~\cite{UPS-prod}. Furthermore, factorization is safer in the 
decay process than in production.  A recent review can be found in
chapter 9 of Ref.~\cite{B-workshop}.

An alternative, not unrelated physical picture begins with a Fock state 
expansion of the $\chi_b$ into the minimal $|b \bar{b} \rangle$ 
color-singlet component and higher components such as 
$|b \bar{b} g\rangle$, with the $b \bar{b}$ pair in a color-octet 
configuration, $|b \bar{b} q \bar{q}\rangle$,  $|b \bar{b} g g\rangle$, 
and so forth.  

A velocity scaling rule in NRQCD~\cite{BBL} allows one to order the 
contributions to the decay rate Eq.~(\ref{eq:factorization})
and to truncate the series on the right hand side of 
Eq.~(\ref{eq:factorization}) in a power series in $v_b$, the 
mean velocity of the $b$ in the $\chi_b$ rest frame.  For  
$\chi_{bJ}$, there are two potentially equally important contributions 
to the decay rate.  
One is the color-singlet spin-triplet $P-$wave state that scales as 
$\langle \chi_{bJ}|%
{\cal O}(^3P_J^{(1)})|\chi_{bJ}\rangle/m_b^2\sim m_b^3v_b^2$.
The other is the color-octet spin-triplet $S-$wave state that is 
proportional to 
$\langle \chi_{bJ}|{\cal O}(^3S_1^{(8)})|\chi_{bJ}\rangle\sim m_b^3v_b^2$, 
where we keep the mass dimension while suppressing the $v_b$ dependence 
of the heavy quark fields and the quarkonium state.  The scaling factor 
$v_b^2$ of the color-octet matrix element arises from the chromoelectric dipole
transition rate from a color-singlet $P$-wave state into a color-octet
spin-triplet state. This color-octet channel was introduced to 
resolve the infrared problem in the next-to-leading order 
QCD corrections to light-hadron decays of $P-$wave quarkonium~\cite{BBL-P}.

In the case of $\Upsilon$ decay to bottom quarks treated in Ref.~\cite{elblc}, 
the leading color-singlet contribution is proportional to $\alpha_s^2v_b^0$.  
The leading contribution in the standard model decay of $\Upsilon$ to light 
hadrons is of order $\alpha_s^3 v_b^0$, with one more power of $\alpha_s$ 
than in the SUSY case.  The SUSY rate is suppressed, however, by the mass 
of the exchanged gluino  in the amplitude and the non-zero bottom 
squark masses.  The leading color-octet contribution in $\Upsilon$ decay is 
of order $\alpha_s^2 v_b^4$.   

In conventional QCD, the $\chi_{bJ}$ states are assumed to decay via the 
transition of the $b \bar{b}$ system into a pair of massless gluons that, in 
turn, materialize as light hadrons~\cite{P-LO}.  The short-distance 
coefficients $C_n$ are known in next-to-leading order in $\alpha_s$ for the 
color-singlet states $n=^3P_J^{(1)}$~\cite{P-NLO,Petrelli} and for the 
color-octet state $n=^3S_1^{(8)}$~\cite{Petrelli,comment}.
We quote the leading-order result in 
$\alpha_s v_b^2$~\cite{positronium,P-LO,BBL-P,BBL}:
\begin{subequations}
\label{eq:chi-SM}
\begin{eqnarray}
\Gamma^{\rm SM}(\chi_{b0})&=&
\frac{4\pi\alpha_s^2 \mathcal{H}_1 }{3m_b^4}
\left(
1
+
\frac{n_f m_b^2\mathcal{H}_8}{4\mathcal{H}_1}
\right),
\\
\Gamma^{\rm SM}(\chi_{b1})&=&
\frac{4\pi\alpha_s^2 \mathcal{H}_1 }{3m_b^4}
\left(
0
+
\frac{n_f m_b^2\mathcal{H}_8}{4\mathcal{H}_1}
\right),
\\
\Gamma^{\rm SM}(\chi_{b2})&=&
\frac{4\pi\alpha_s^2 \mathcal{H}_1 }{3m_b^4}
\left(
\frac{4}{15}
+
\frac{n_f m_b^2\mathcal{H}_8}{4\mathcal{H}_1}
\right).
\end{eqnarray}
\end{subequations}
The superscript $\rm SM$ designates the standard model contribution 
without SUSY terms.  The number of flavors $n_f = 4$.  
The matrix elements in Eq.~(\ref{eq:chi-SM}) are the color-singlet 
term $\mathcal{H}_1=\langle h_b|\mathcal{O}(^1P_1^{(1)})| h_b\rangle$
and the color-octet term 
$\mathcal{H}_8=\langle h_b|\mathcal{O}(^1S_0^{(8)})| h_b\rangle$; 
$h_b$ is the spin-singlet $P-$wave bottomonium state.
Heavy quark spin symmetry is used in the approximations 
$\langle \chi_{bJ}|\mathcal{O}(^3P_J^{(1)})|\chi_{bJ}\rangle=\mathcal{H}_1%
+O(v_b^2)$, and
$\langle \chi_{bJ}|\mathcal{O}(^3S_1^{(8)})|\chi_{bJ}\rangle=\mathcal{H}_8%
+O(v_b^2)$.
The color-singlet matrix element can be expressed in terms of the derivative 
of the radial wave function at the origin, 
$\mathcal{H}_1=\frac{3N_c}{2\pi}|R'(0)|^2 + O(v_b^2)$.
Numerical values of these matrix elements are available from lattice
measurements~\cite{BKS1,BKS2}:
\begin{eqnarray} 
\mathcal{H}_1 =2.7~\textrm{GeV}^5,\quad
\mathcal{H}_8/\mathcal{H}_1
       =2.275\times 10^{-3}~\textrm{GeV}^{-2}.
\label{eq:H18}
\end{eqnarray}
As is evident in Eq.~(\ref{eq:chi-SM}b), the decay rate for the $\chi_{b1}$ is 
purely color octet in nature at leading order in $\alpha_s$ because the transition 
of the $J^{PC} = 1^{++}$ color-singlet state into two massless gluons is forbidden.  
Since $v_b^2 \sim 0.1$, the color-octet matrix element $\mathcal{H}_8$ is small 
for bottomonium (for charmonium $v_c^2 \sim 0.3$ is not negligible). The 
color-octet contributions in Eq.~(\ref{eq:chi-SM}) are estimated to be at the 
5\% level in the $\chi_{b0}$ case and at 
the 20\% level in the $\chi_{b2}$ case.  The numerical values of the predicted SM 
hadronic widths in Eq.~(\ref{eq:chi-SM}) are approximately 0.8 MeV, 0.04 MeV, and 
0.2 MeV for the $\chi_{b0}$, $\chi_{b1}$, and $\chi_{b2}$, respectively.  

\subsection{Derivation of the Short-Distance Term}

A practical procedure to extract the short-distance coefficient $C_n$ 
in Eq.~(\ref{eq:factorization}) begins from a calculation of 
the free $b\overline{b}$ amplitude with specified spin, color, and
orbital angular momentum. 
We use $P$ and $q$ to denote the total and the relative momenta of the
pair.  The heavy quark momentum $p_b$ and its counterpart $p_{\overline{b}}$ 
are 
\begin{eqnarray}
p_b=\frac{P}{2}+q,\quad
p_{\overline{b}}=\frac{P}{2}-q.
\end{eqnarray}
In the rest frame of the $b\overline{b}$ pair, the components 
of the vectors become $P=(2E_q,\mathbf{0})$,
$q=(0,\mathbf{q})$, $p_b=(E_q,\mathbf{q})$, and
$p_{\overline{b}}=(E_q,-\mathbf{q})$.  
In the non-relativistic limit, the invariant mass of the pair
can be expanded as $2E_q=2\sqrt{m_b^2+\mathbf{q}^2}=2m_b+O(\mathbf{q}^2)$.

The amplitude for transition of the free $b\overline{b}$ system into a final 
state is denoted $\mathcal{M}_{b\overline{b}}$. It includes a spinor 
factor $u^i_\alpha(p_b,s)\bar{v}^j_\beta(p_{\overline{b}},\bar{s})$
where $(s,\alpha,i)$ and $(\bar{s},\beta,j)$ are the 
(spin, spinor index, color index) of the $b$ and $\overline{b}$, respectively.
The spin-triplet state of the pair can be projected from the outer product
of the heavy quark spinors upon multiplying with Clebsch-Gordan 
coefficients~\cite{Salpeter,P-LO,Berger:1980ni,BL}.
\begin{subequations}
\label{eq:projection}
\begin{eqnarray}
&&\mathcal{M}_{b\overline{b}}=
\mathcal{M}^{ji}_{\beta\alpha}(p_b,p_{\overline{b}})\;
\sum_{s,\bar{s}}
\langle 1\lambda|s,\bar{s}\rangle
u^i_\alpha(p_b,s)\bar{v}^j_\beta(p_{\overline{b}},\bar{s}),
\label{eq:amplitude-QQ}
\\
&&\sum_{s,\bar{s}}
\langle 1\lambda|s,\bar{s}\rangle 
u^i_\alpha(p_b,s)\bar{v}^j_\beta(p_{\overline{b}},\bar{s})
=
\Lambda_{\alpha\beta}^{ijk}
\langle 0|\chi^\dagger\sigma^k\psi| b\overline{b}\rangle
+
\Lambda_{\alpha\beta}^{aijk}
\langle 0|\chi^\dagger T^a\sigma^k\psi| b\overline{b}\rangle,
\label{eq:projection-all}
\\
&&\Lambda_{\alpha\beta}^{ijk}
=
\frac{\delta^{ij}}{N_c}\times
\frac{1}{4m_b}\left[
              \left(\not{p_b}+m_b\right)\gamma_\mu
              \left(\not{p_{\overline{b}}}-m_b\right)\right]_{\alpha\beta} {L^\mu}_k,
\label{eq:projection-1}
\\
&&\Lambda_{\alpha\beta}^{aijk}
=
2T^a_{ij}\times
\frac{1}{4m_b}\left[
              \left(\not{p_b}+m_b\right)\gamma_\mu
              \left(\not{p_{\overline{b}}}-m_b\right)\right]_{\alpha\beta} {L^\mu}_k,
\label{eq:projection-8}
\end{eqnarray}
\end{subequations}
where ${L^\mu}_k$ is the boost matrix from the $b\overline{b}$ rest 
frame to the frame in which the $b\overline{b}$ pair has momentum 
$P$.  The operators $\psi$ and  $\chi^\dagger$ are the
annihilation operators for the non-relativistic 
$b$ and $\overline{b}$ states, respectively.
The projection operator Eq.~(\ref{eq:projection}) is valid up to terms 
of $O(\textbf{q}^2)$. The projection operator valid to 
all orders in $\textbf{q}^n$ is derived in Ref.~\cite{BL}.
The heavy quark state in Eq.~(\ref{eq:projection-all})
is specified with the non-relativistic normalization
$\langle b(\textbf{p})| b(\textbf{q})\rangle%
=(2\pi)^3\delta^{(3)}(\textbf{p}-\textbf{q})$, whereas the Dirac spinors 
have relativistic normalization, $\bar{u} u=\bar{v}v=2m_b$.

Substituting the projection operator Eq.~(\ref{eq:projection-all})
into Eq.~(\ref{eq:amplitude-QQ}) and expanding with respect to
$\textbf{q}$ up to $O(\textbf{q}^2)$, we can express
the amplitude $\mathcal{M}_{b\overline{b}}$ as a linear combination
of a color-singlet $P-$wave term and a color-octet $S-$wave term.  
\begin{eqnarray}
\mathcal{M}_{b\overline{b}}&=&
\mathcal{A}^{ij}\langle 0|\chi^\dagger \textbf{q}^i\sigma^j\psi| 
b\overline{b}\rangle
+\mathcal{B}^{ai} \langle 0|\chi^\dagger T^a\sigma^i\psi| 
b\overline{b}\rangle.
\label{eq:amplitude-PS}
\end{eqnarray}
The $P$-wave amplitude can be decomposed further into
the total angular momentum components, $J=0,1,2$.  
\begin{eqnarray}
\mathcal{M}_{b\overline{b}}&=& 
\mathcal{A}_0 
\langle 0|\chi^\dagger \mathcal{K}(^3P_0^{(1)})\psi|b\overline{b}\rangle
+\mathcal{A}_1^i
\langle 0|\chi^\dagger \mathcal{K}^i(^3P_1^{(1)})\psi|b\overline{b}\rangle
\nonumber\\ &+&
\mathcal{A}_2^{ij}
\langle 0|\chi^\dagger \mathcal{K}^{ij}(^3P_2^{(1)})\psi|b\overline{b}\rangle
+\mathcal{B}^{ai}
\langle 0|\chi^\dagger \mathcal{K}^{ai}(^3S_1^{(8)})\psi|
b\overline{b}\rangle.  
\label{eq:amplitude-PJ}
\end{eqnarray}
The coefficients in Eq.~(\ref{eq:amplitude-PJ}) are 
$\mathcal{A}_0=\mathcal{A}^{ii}/\sqrt{3}$,
$\mathcal{A}_1^i=\epsilon^{ijk}\mathcal{A}^{jk}/\sqrt{2}$, and
$\mathcal{A}_2^{ij}=({A}^{ij}+{A}^{ji})/2-\delta^{ij}\mathcal{A}^{kk}/3$.
The corresponding operators are 
$\mathcal{K}(^3P_0)=\textbf{q}\cdot\bm{\sigma}/{\sqrt{3}}$,
$\mathcal{K}^i(^3P_1)=(\textbf{q}\times\bm{\sigma})^i/{\sqrt{2}}$, and
$\mathcal{K}^{ij}(^3P_2)=
(\textbf{q}^i\sigma^j+\textbf{q}^j\sigma^i)/2
                    -\delta^{ij}\bm{\sigma}\cdot\textbf{q}/3$.

The squared and spin-averaged amplitude for the $^3P_J$ state is 
\begin{eqnarray}
\overline{|\mathcal{M}_{b\overline{b}}|^2}&=&
\sum_{J=0}^{2}
|\mathcal{A}_J|^2
\langle b\overline{b}|\mathcal{O}(^3P_J^{(1)})|b\overline{b}\rangle
+
|\mathcal{B}|^2
\langle b\overline{b}|\mathcal{O}(^3S_1^{(8)})|b\overline{b}\rangle.  
\label{eq:amplitude2-PJ}
\end{eqnarray}
The coefficients $|\mathcal{A}_J|^2$ and
$|\mathcal{B}|^2$ are obtained by contracting to
rotationally symmetric tensors:
\begin{subequations}
\label{eq:pol-sum}
\begin{eqnarray}
|\mathcal{A}_1|^2&=&\frac{\delta^{ij}}{3}
                             \mathcal{A}_1^i
                             \mathcal{A}_1^{j*},
\\
|\mathcal{A}_2|^2&=&
\frac{1}{5}
\left[
\frac{1}{2}\left( \delta^{ik}\delta^{jl}+\delta^{il}\delta^{jk}\right)
-\frac{1}{3}\delta^{ij}\delta^{kl}
\right]
                             \mathcal{A}_2^{ij}
                             \mathcal{A}_2^{kl*},
\\
|\mathcal{B}|^2&=&\frac{\delta^{ij}}{3}
                  \mathcal{B}^{i}
                  \mathcal{B}^{j*}.
\end{eqnarray}
\end{subequations}
The coefficient $\mathcal{B}^{ai}$ of the color-octet term
in the amplitude Eq.~(\ref{eq:amplitude-PJ})
has a free color index.
However, owing to color conservation, the squared term has the form 
$\mathcal{B}^{ai}\mathcal{B}^{*bj}=\delta^{ab}\mathcal{B}^{i}\mathcal{B}^{*j}$,  
and the color index of the color-octet matrix element is contracted
with the factor $\delta^{ab}$.

The last step to obtain the short distance coefficients is 
to introduce the normalization factor
$(2M_{\chi_{QJ}})/(2E_q)^2=1/m_Q+O(\textbf{q}^2)$.
This factor appears since we use non-relativistic normalization
in the matrix elements for the $b\overline{b}$ and $\chi_{bJ}$ states.
The short-distance coefficients are therefore expressed as
\begin{eqnarray}
C(^3P_J^{(1)})=\frac{\Phi}{4m^2_b}|\mathcal{A}_J|^2,\quad
C(^3S_1^{(8)})=\frac{\Phi}{4m^2_b}|\mathcal{B}|^2.  
\label{eq:short-PJ1}
\end{eqnarray}
The non-relativistic zero-binding energy approximation $m_{\chi_{bJ}}=2m_b$ 
is used.  For a final state composed of two particles with identical mass $m_f$, 
the phase space factor $\Phi=\frac{1}{8\pi} (1-m_f^2/m_b^2)^{1/2}$.  

\section{Decay into Bottom Squarks\label{decaystosbottoms}}

In this section we treat the inclusive decay of a $\chi_b$ into a pair of 
bottom squarks $\widetilde{b}$ 
of mass $m_{\tilde{b}}$ carrying four-momenta $k_1$ and $k_2$ respectively.  The  
relevant short-distance perturbative subprocess is 
\begin{equation}
b + \bar{b} \rightarrow \widetilde{b} + {\widetilde{b}^*} .
\label{reaction}
\end{equation}
The Feynman diagrams for this subprocess are shown in 
Fig.~\ref{fig:feynman}.  In Fig.~\ref{fig:feynman}(a), a $t$-channel gluino 
$\widetilde{g}$ of mass $m_{\tilde{g}}$ is exchanged.   
The subprocess sketched in Fig.~\ref{fig:feynman}(b) in which the $b \bar{b}$ 
pair annihilates through an intermediate gluon into a $\tilde{b} {\tilde{b}}^*$
pair is absent in the color-singlet approximation because the initial state is a 
color singlet.  Both Fig.~\ref{fig:feynman}(a) and Fig.~\ref{fig:feynman}(b) 
contribute to the color-octet amplitude.  Figure~\ref{fig:feynman}(b) is 
independent of the gluino mass, and its contribution survives therefore even 
in the limit of gluinos with very large mass.  Figure~\ref{fig:feynman}(a) contributes 
to both the $^3S_1^{(1)}$ 
and $^3P_J^{(1)}$ channels whereas the analogous SM process in which gluons are 
produced and a $b$ quark is exchanged contributes only to $^3P_{0,2}^{(1)}$.

\begin{figure}
\includegraphics[height=3.5cm]{fig1.epsi}
\caption{Feynman diagrams for the process
$b\overline{b}\to \tilde{b}_{1} {\tilde{b}}^*_{1}$. }
\label{fig:feynman}
\end{figure}

\subsection{Bottom Squark Mixing and Couplings\label{sbottoms}}

The mass eigenstates of the bottom squarks, $\tilde{b}_1$ and $\tilde{b}_2$ 
are mixtures of left-handed (L) and right-handed (R) bottom squarks, 
$\tilde{b}_L$ and $\tilde{b}_R$.  The mixing is expressed as 
\begin{eqnarray}
|\tilde{b}_1\rangle = \sin\theta_{\tilde{b}}|\tilde{b}_L\rangle + 
				\cos\theta_{\tilde{b}}|\tilde{b}_R\rangle, \\
|\tilde{b}_2\rangle = \cos\theta_{\tilde{b}}|\tilde{b}_L\rangle - 
				\sin\theta_{\tilde{b}}|\tilde{b}_R\rangle.   
\label{eq:mixing}
\end{eqnarray}
In our notation, the lighter mass eigenstate is denoted $\tilde{b}_1$.
For the case under consideration, the mixing of the bottom squark is 
determined by the condition that the coupling of the lighter $\tilde b$ 
to the $Z$ boson be small~\cite{CHWW}, 
namely $\sin^2\theta_{\tilde{b}} \simeq 1/6$.

We may also express the mass eigenstate $\tilde{b}_1$ in terms of states of 
definite parity, the $J^P = 0^+$ scalar and $J^P = 0^-$ pseudo-scalar.  
Starting from the relationships 
\begin{eqnarray}
|0^+\rangle = \frac{1}{\sqrt{2}}(|\tilde{b}_R\rangle + |\tilde{b}_L\rangle), \\
|0^-\rangle = \frac{1}{\sqrt{2}}(|\tilde{b}_R\rangle - |\tilde{b}_L\rangle), 
\label{eq:parity1}
\end{eqnarray}
we obtain    
\begin{equation}
|\tilde{b}_1\rangle = 
\frac{1}{\sqrt{2}}(\cos\theta_{\tilde{b}} + \sin\theta_{\tilde{b}})|0^+\rangle + 
\frac{1}{\sqrt{2}}(\cos\theta_{\tilde{b}} - \sin\theta_{\tilde{b}})|0^-\rangle. 
\label{eq:parity2}
\end{equation}
The $\tilde{b}_1$ is a pure $J^P = 0^+$ scalar only if 
$\sin\theta_{\tilde{b}} = \cos\theta_{\tilde{b}} = \frac{1}{\sqrt{2}}$. 

The coupling at the three-point vertex in which a $b$ quark enters and a 
$\tilde{b}_1$ squark emerges (the upper vertex in Fig.\ref{fig:feynman}(a)) is 
\begin{equation}
i g_s \sqrt 2 T^a_{ki} [\cos\theta_{\tilde{b}}P_R - \sin\theta_{\tilde{b}}P_L],
\label{upper}
\end{equation}
where $i$ and $k$ are the color indices of the incident $b$ and final 
$\tilde{b}$, respectively, and $a$ labels the color of the exchanged 
gluino.  Here, $P_L = (1 - \gamma_5)/2$ and $P_R = (1 + \gamma_5)/2$.  At the 
lower vertex where an antiquark enters and ${\tilde{b}}_1^*$ emerges, the coupling 
is 
\begin{equation}
i g_s \sqrt 2 T^a_{j\ell} [\cos\theta_{\tilde{b}}P_L - \sin\theta_{\tilde{b}}P_R],
\label{lower}
\end{equation}
where $j$ and $\ell$ are the color indices of the incident $\bar{b}$ and 
final ${\tilde{b}}^*$, respectively.  

\subsection{Subprocess Amplitude\label{subamplitude}}

The amplitude of the process 
$b\overline{b}\to \tilde{b}_{\lambda} {\tilde{b}}^*_{\bar{\lambda}}$,
where $\lambda(\bar{\lambda})=-1$ for $\tilde{b}_L({\tilde{b}}^*_L)$ and 
$\lambda=1$ for $\tilde{b}_R({\tilde{b}}^*_R)$, respectively, is 
\begin{eqnarray}
\mathcal{M}^{ji}_{\beta\alpha}(\lambda,\bar{\lambda})
&=&
\left[
\frac{ig_s\bar{\lambda}}{\sqrt{2}}T^a_{jl}\left(1-\bar{\lambda}\gamma_5\right)
\times
\frac{i}{\not{p}_b-\not{k_1}-m_{\tilde{g}}}
\times
\frac{ig_s\lambda}{\sqrt{2}}T^a_{ki}\left(1+\lambda\gamma_5\right)
\right]_{\beta\alpha}
\nonumber\\
&+&
\left(-ig_sT^a_{ji}\gamma^\mu\right)_{\beta\alpha}
\times
\frac{-ig_{\mu\nu}}{P^2}
\times
\left[-ig_sT^a_{kl}(k_1-k_2)^\nu\right].
\label{eq:amplitude-sb}
\end{eqnarray}
Momentum labels are specified in Fig.\ref{fig:feynman}; $P$ is the 
four-momentum of the $\chi_b$.  As remarked earlier, 
$\alpha$ and $\beta$ are the spinor indices of the $b$ and 
$\overline{b}$.  The first term in Eq.~(\ref{eq:amplitude-sb})
represents the diagram shown in Fig.~\ref{fig:feynman}(a), and the other term is
for the diagram of Fig.~\ref{fig:feynman}(b).
We do not simplify the factors in Eq.~(\ref{eq:amplitude-sb})
so that the Feynman rules can be identified in this expression. 
The relatively large masses of the exchanged gluino Fig.\ref{fig:feynman}(a) and 
of the $\chi_b$ should justify the use of simple perturbation theory to 
compute the decay amplitude;  $g_s$ is the strong coupling, $\alpha_s = g_s^2/4\pi$.
  
Using the mixing relation Eq.~(\ref{eq:mixing}), we can display the explicit mixing 
angle dependence of the amplitude for the light bottom squark pair final state in 
the form 
$\mathcal{M}^{ji}_{\beta\alpha}%
=%
\sin^2\theta\mathcal{M}^{ji}_{\beta\alpha}(+,+)%
+\cos^2\theta\mathcal{M}^{ji}_{\beta\alpha}(-,-)%
+\sin\theta\cos\theta(%
\mathcal{M}^{ji}_{\beta\alpha}(+,-)%
+\mathcal{M}^{ji}_{\beta\alpha}(-,+))$.

Substituting $\mathcal{M}^{ji}_{\beta\alpha}$
into Eq.~(\ref{eq:amplitude-QQ}) and using the projection operator
Eq.~(\ref{eq:projection-all}), we derive an expression in the form
of Eq.~(\ref{eq:amplitude-PS}).
Useful color factor formulas are
\begin{subequations}
\begin{eqnarray}
\left[T^a_{ki}T^a_{jl}\right]\delta_{ij}&=&\frac{N_c^2-1}{2N_c}\delta_{kl},
\\
\left[T^a_{ki}T^a_{jl}\right]T^b_{ij}&=&-\frac{1}{2N_c}T^b_{kl}.
\end{eqnarray}
\end{subequations}
The partial-wave amplitudes in Eq. $(\ref{eq:amplitude-PJ})$ are found to be
\begin{subequations}
\begin{eqnarray}
\mathcal{A}_0&=&
\frac{32\pi\alpha_s}{9\sqrt{3}}\delta_{kl}
  \frac{6\sin\theta_{\tilde{b}}\cos\theta_{\tilde{b}} m_{\tilde{g}}
                  (m_b^2+m_{\tilde{g}}^2-m_{\tilde{b}}^2)
                 -m_b(m_b^2+3m_{\tilde{g}}^2-m_{\tilde{b}}^2)
       }
       {(m_b^2+m_{\tilde{g}}^2-m_{\tilde{b}}^2)^2},
\\
\mathcal{A}^{i}_1&=&
\frac{64\pi\alpha_s}{9\sqrt{2}}\delta_{kl} 
\frac{k_{1\mu}{L^\mu}_i (\cos^2\theta_{\tilde{b}}-\sin^2\theta_{\tilde{b}})}
     {(m_b^2+m_{\tilde{g}}^2-m_{\tilde{b}}^2)},
\\
\mathcal{A}^{ij}_2&=&
\frac{64\pi\alpha_s}{9}\delta_{kl} \frac{m_b
                      k_{1\mu} k_{1\nu} {L^\mu}_i{L^\nu}_j}
                    {(m_b^2+m_{\tilde{g}}^2-m_{\tilde{b}}^2)^2},
\\
\mathcal{B}^{ai}&=&\frac{4\pi\alpha_s}{3} T^a_{kl} \frac {k_{1\mu}{L^\mu}_i}{m_b}
 \left( \frac {2m_b^2}{m_b^2+m_{\tilde{g}}^2-m_{\tilde{b}}^2} + 3 \right) .
\label{eq:PWB}
\end{eqnarray}
\end{subequations}
Using the polarization summation relations in Eq.~(\ref{eq:pol-sum}),
we obtain the short-distance factors of Eq.~(\ref{eq:short-PJ1}).
Substituting these short-distance coefficients into the factorization 
formula Eq.~(\ref{eq:factorization}), we derive  
\begin{subequations}
\begin{eqnarray}
\Gamma(\eta_b\to \tilde{b}_1 {\tilde{b}}^*_1)&=&
0, 
\label{eq:Geta}
\\
\Gamma(\Upsilon\to \tilde{b}_1 {\tilde{b}}^*_1)&=&
\frac{32\pi\alpha_s^2\mathcal{G}_1}{81}
 \left(1-\frac{m_{\tilde{b}}^2}{m_b^2}\right)^{3/2}
 \frac{m_b^2}{(m_b^2+m_{\tilde{g}}^2-m_{\tilde{b}}^2)^2},
\label{eq:GUpsilon}
\\
\Gamma(\chi_{bJ}\to \tilde{b}_1 {\tilde{b}}^*_1)&=&
\frac{4\pi\alpha_s^2 \mathcal{H}_1}{3m_b^4}
\left(1-\frac{m_{\tilde{b}}^2}{m_b^2}\right)^{1/2}
\left( D_J+D_8\frac{m_b^2\mathcal{H}_8}{\mathcal{H}_1}\right). 
\label{eq:GChi} 
\end{eqnarray}
\end{subequations}
where 
$\mathcal{G}_1 = 
\langle \Upsilon|{\cal O}(^3S_1^{(1)})|\Upsilon\rangle\ = 
\frac{N_c}{2\pi}|R(0)|^2$, and $R(0)$ is the $\Upsilon$ wave function at the 
origin.  

The coefficients are 
\begin{subequations}
\begin{eqnarray}
D_0&=&
\frac{8}{27}
 \frac{m_b^2\left[
       6m_{\tilde{g}}(m_b^2+m_{\tilde{g}}^2-m_{\tilde{b}}^2)
         \sin\theta_{\tilde{b}}\cos\theta_{\tilde{b}}
       -m_b(m_b^2+3m_{\tilde{g}}^2-m_{\tilde{b}}^2)\right]^2
}
      {(m_b^2+m_{\tilde{g}}^2-m_{\tilde{b}}^2)^4},
\label{eq:coefD0}
\\
D_1&=&
\frac{16}{27}\left(1-\frac{m_{\tilde{b}}^2}{m_b^2}\right)
 \frac{
 m_b^4
\left(\cos^2\theta_{\tilde{b}}-\sin^2\theta_{\tilde{b}}\right)^2
       }{(m_b^2+m_{\tilde{g}}^2-m_{\tilde{b}}^2)^2},
\label{eq:coefD1}
\\
D_2&=&
\frac{64}{135}
 \left(1-\frac{m_{\tilde{b}}^2}{m_b^2}\right)^{2}
 \frac{m_b^8}{(m_b^2+m_{\tilde{g}}^2-m_{\tilde{b}}^2)^4},
\label{eq:coefD2}
\\
D_8&=&
\frac{1}{144}
 \left(1-\frac{m_{\tilde{b}}^2}{m_b^2}\right)
 \left ( 3 + \frac{2m_b^2}{m_b^2+m_{\tilde{g}}^2-m_{\tilde{b}}^2} \right)^2 .
\label{eq:coefD8}
\end{eqnarray}
\end{subequations}
Equation~(\ref{eq:GUpsilon}) was derived previously in Ref.~\cite{elblc}. 
Equation~(\ref{eq:Geta}) indicates that $\eta_b$ decay is not allowed in 
lowest-order.  In the color-singlet case, reduction of the relevant trace 
provides a lowest-order amplitude proportional to 
$\left(\cos^2\theta_{\tilde{b}} - \sin^2\theta_{\tilde{b}} \right) 
p_{\eta} \cdot (p_{\eta} - 2 k_1)$, where $p_{\eta}$ and $k_1$ are the 
four-vector momenta of the $\eta_b$ and one of the final $\tilde{b}$'s.   
Evaluation in the $\eta_b$ rest frame shows that this amplitude is 
zero for a two-body final state.  

A symmetric S-wave two-body $\widetilde{q} \widetilde{q}^*$ system can be 
constructed with one of the pair in a $J^P = 0^-$ state and the other a 
$J^P = 0^+$ state.  Under charge-conjugation $C$, a squark $\widetilde{q}$ is 
transformed into an anti-squark $\widetilde{q}^*$ with no overall phase.  
The two-body $\widetilde{q} \widetilde{q}^*$ system will have $J^{PC} = O^{-+}$, 
the same quantum numbers as the $\eta_b$.  Therfore, there appears to be no overall 
symmetry that would forbid the exclusive lowest-order process 
$\eta_b \rightarrow \tilde{b}_1 \tilde{b}^*_1$, and Eq.~(\ref{eq:Geta}) is not 
true more generally than in the lowest-order model we use.  

As remarked above, the pure color-octet Fig.~\ref{fig:feynman}(b) term is 
independent of the gluino mass, and its contribution survives therefore even in 
the limit of gluinos with very large mass.  Equation~(\ref{eq:PWB}) is written 
in a form that makes this fact transparent; the solitary $``$3" represents the 
contribution in the limit that only Fig.~\ref{fig:feynman}(b) contributes.  
If $m^2_{\tilde{b}} \ll m^2_{\tilde{g}}$ and $m^2_b \ll m^2_{\tilde{g}}$, 
the color-singlet coefficients $D_J$, Eqs.~(\ref{eq:coefD0} - \ref{eq:coefD2}), 
vanish in proportion to $m_{\tilde{g}}^{-n}$, $n = 2^{J + 1}$, but the 
color-octet coefficient $D_8$ remains finite.  
As $m_{\tilde{g}} \rightarrow \infty$, 
\begin{equation}
\Gamma(\chi_{bJ}\to \tilde{b}_1 {\tilde{b}}^*_1) \rightarrow 
\frac{4\pi\alpha_s^2 \mathcal{H}_8}{3m_b^2} \frac{1}{16}
\left(1-\frac{m_{\tilde{b}}^2}{m_b^2}\right)^{3/2} = 
\Gamma^{\rm SM}(\chi_{b1})\frac{1}{4n_f}\left(1-\frac{m_{\tilde{b}}^2}{m_b^2}\right)^{3/2} .
\label{eq:GChiinf} 
\end{equation}

The momentum of a final state bottom squark in the rest frame of the 
$\chi_b$ is $|k| = \frac{1}{2}\sqrt {m^2_{\chi_b} - 4m^2_{\tilde{b}}} \simeq 
m_b\sqrt{1-\frac{m_{\tilde{b}}^2}{m_b^2}}$.  
We observe that $\Gamma  \propto |k|^{2\ell +1}$, with $\ell = 1$, in the case 
of $\Upsilon$ decay, Eq.~(\ref{eq:GUpsilon}), as expected because the decay of 
the $\Upsilon$ produces a pair of scalars in a $P$-wave.  In the $\chi_J$ case, 
combining the overall phase space factor $\propto |k|$ in Eq.~(\ref{eq:GChi}), with 
those $\propto |k|^{2J}$ in the expressions for $D_J$ in 
Eqs.~(\ref{eq:coefD0} - \ref{eq:coefD2}), we note that the color-singlet terms 
have the threshold behaviors expected for decays into $S, P, {\rm and}~D$ wave 
systems of two spin-0 particles.   The overall color-octet coefficient 
Eqs.~(\ref{eq:GChi}) and ~(\ref{eq:coefD8}) is proportional to $|k|^3$.  

The dependences on the mixing angle $\theta_{\tilde{b}}$ in 
Eqs.~(\ref{eq:coefD0} - \ref{eq:coefD8}) can be understood if we recast the 
results in terms of production of left-handed and right-handed bottom squarks, 
$\tilde{b}_L$ and $\tilde{b}_R$.  In the $\chi_{b0}$ case, the left-left 
and right-right combinations are produced with equal rates, leading to a 
term in Eq.~(\ref{eq:coefD0}) proportional to 
$(\cos^2\theta_{\tilde{b}} + \sin^2\theta_{\tilde{b}}) = 1$. In addition, 
the left-right and right-left combinations are produced with equal rates, 
leading to the term proportional to 
$\sin\theta_{\tilde{b}}\cos\theta_{\tilde{b}}$.  For $\chi_{b1}$, the left-left 
and right-right combinations are produced with equal but opposite rates, leading 
to a dependence of $(\cos^2\theta_{\tilde{b}} - \sin^2\theta_{\tilde{b}})$.
The left-right and right-left combinations do not contribute.  In the 
$\chi_{b2}$ case, the the left-left and right-right combinations are produced with 
equal rates, resulting in a dependence of $(\cos^2\theta_{\tilde{b}} + 
\sin^2\theta_{\tilde{b}}$), and the left-right and right-left combinations 
do not contribute.

We can obtain a rough estimate of the relative size of the color-octet and 
color-singlet contributions in Eq.~(\ref{eq:GChi}) by beginning with the 
simplifications $(m_b^2 - m^2_{\tilde{b}}) \ll m^2_{\tilde{g}}$ and 
$m^2_b \ll m^2_{\tilde{g}}$, both reasonably fair approximations for the 
masses of interest.  Doing so, and defining $x =m^2_b/m^2_{\tilde{g}}$, 
we find $D_8 : D_0 : D_1 : D_2 
\simeq 1 : 512 x \cos^2\theta_{\tilde{b}}\sin^2\theta_{\tilde{b}}/3 : 
256 x^2 (\cos^2\theta_{\tilde{b}} -\sin^2\theta_{\tilde{b}})^2/27 : 
1024 x^4/135$.  
Adopting the lattice estimate 
$m^2_b\mathcal{H}_8/\mathcal{H}_1 \simeq 0.05$ 
(c.f., Eq.~(\ref{eq:H18})) along with the typical value $x \simeq 1/10$ for 
the range of sparticle masses under consideration, we conclude that the 
color-octet contributions to bottom squark pair production should be roughly 
a 2\% effect for the $\chi_{b0}$, approximately 50\% of the answer for 
$\chi_{b1}$, and dominant in the $\chi_{b2}$ case.  Nevertheless, as shown in 
the next section, bottom squark decays are expected to make a small overall 
contribution to the hadronic widths of the $\chi_{b1}$ and $\chi_{b2}$.  


\section{Predictions and discussion\label{sec:predict}}
\begin{figure}
\includegraphics[height=19cm]{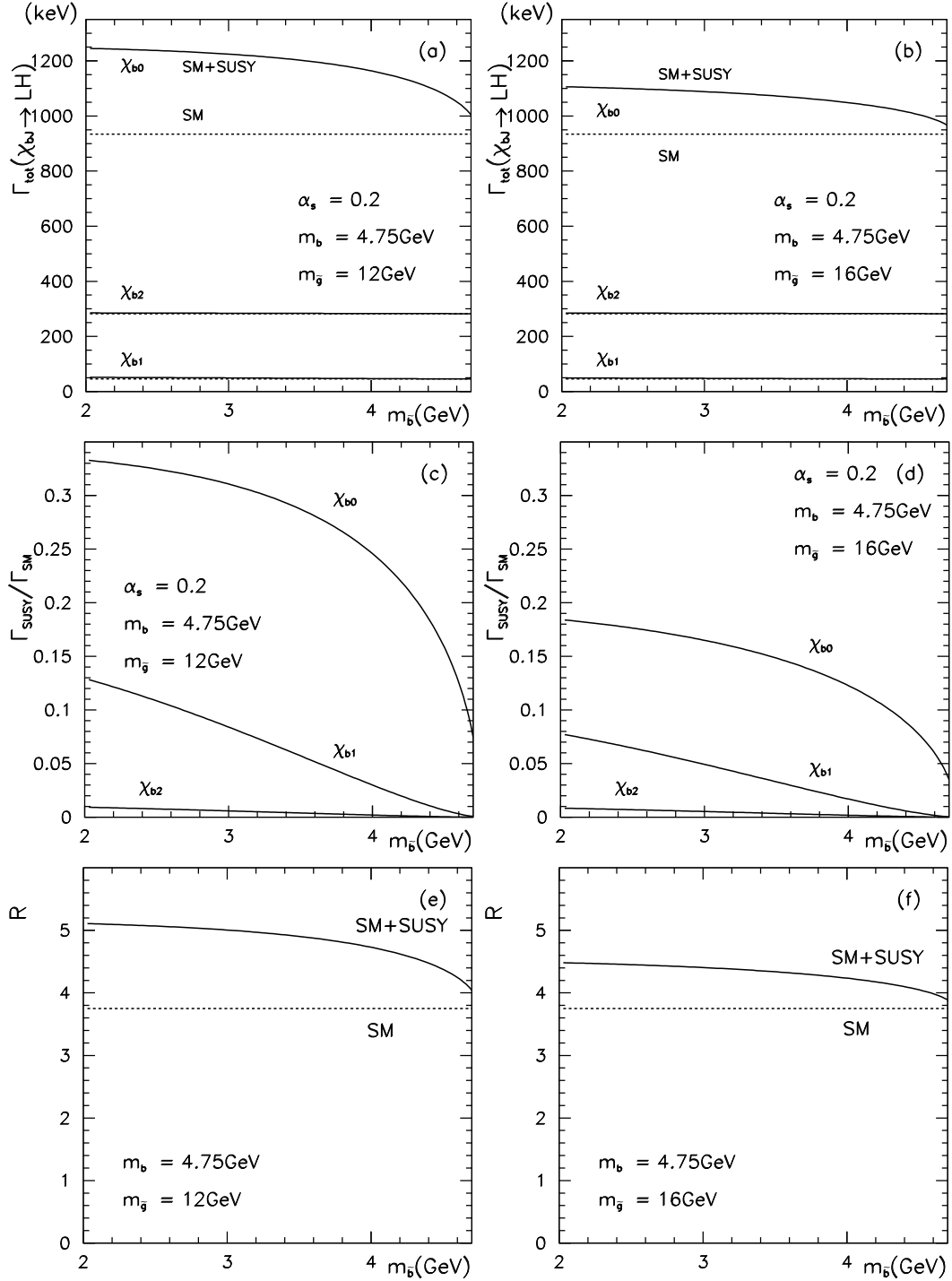}
\caption{Hadronic decay rates of the $\chi_b$ states as functions of the
bottom squark mass are shown in the first row for
$\sin\theta_{\tilde{b}}\cos\theta_{\tilde{b}} > 0$.
The left-hand column shows results for $m_{\tilde{g}} = 12$ GeV and the
right-hand column for $m_{\tilde{g}} = 16$ GeV.  Ratios of the bottom 
squark
decay rates to the standard model predictions are shown in the second row.
The third row shows the ratio $R$ defined in Eq.~(\ref{eq:ratio}).}
\label{figure:2}
\end{figure}

\begin{figure}
\includegraphics[height=19cm]{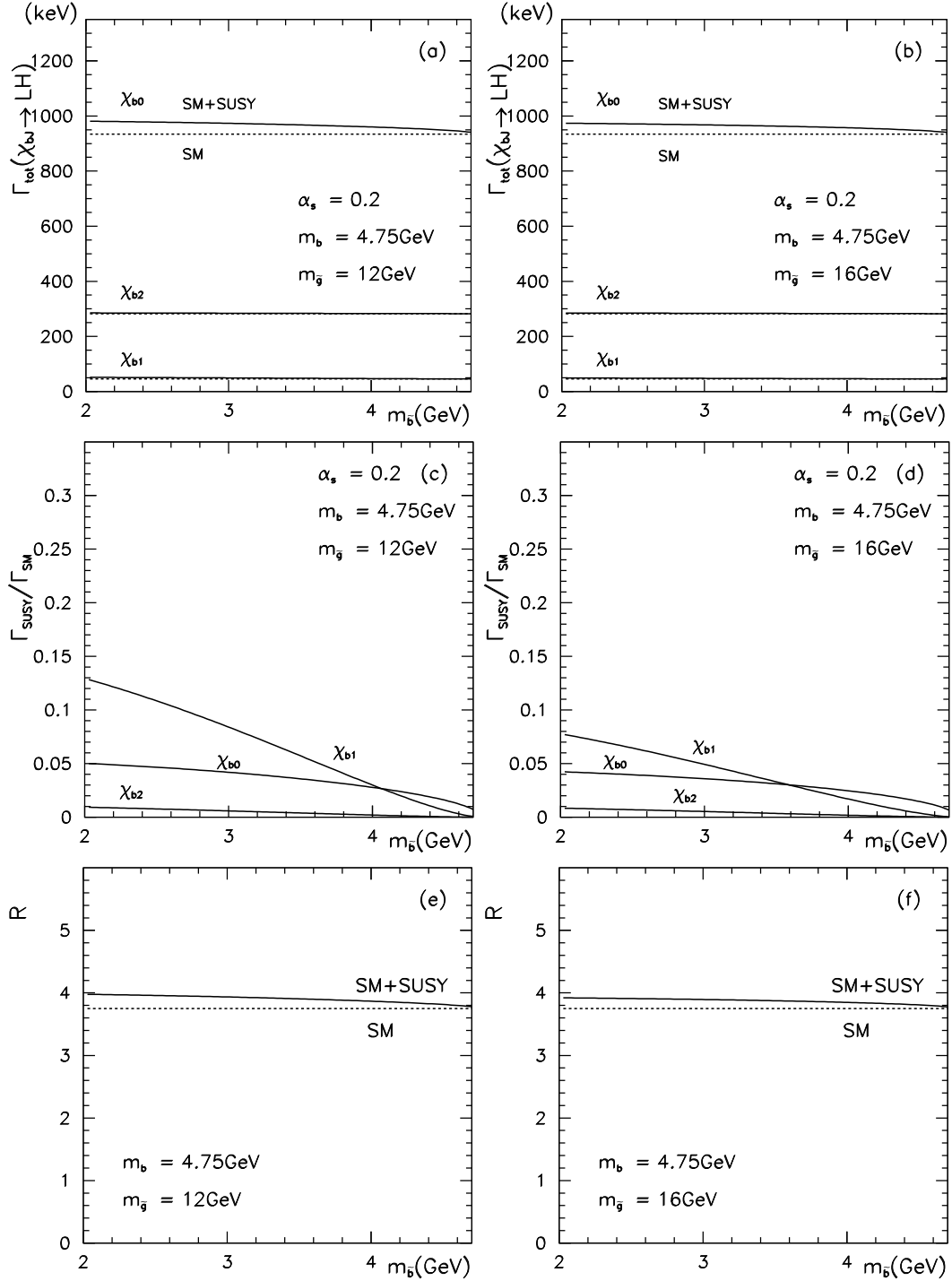}
\caption{As in Fig.~\ref{figure:2}
but with $\sin\theta_{\tilde{b}}\cos\theta_{\tilde{b}} < 0$.}
\label{figure:3}
\end{figure}

The total hadronic decay rates of the $\chi_{bJ}$ states are obtained after 
adding the contributions from conventional two-gluon decay and those for 
bottom squark decay, Eqs.~(\ref{eq:chi-SM}) and (\ref{eq:GChi}).  These final 
decay rates are shown in the first rows of Figs.~\ref{figure:2} and \ref{figure:3} 
and are compared with the standard model expectations.    
The left-hand sides of the figures show results for the choice $m_{\tilde{g}} = 
12$ GeV, and the right-hand sides for $m_{\tilde{g}} = 16$ GeV.  Results are displayed 
as a function of the bottom squark mass in the range 
2~$\textrm{GeV}<m_{\tilde{b}}<m_b$.  The choices of $m_{\tilde{g}}$ and 
$m_{\tilde{b}}$ are guided by the results in Ref.~\cite{BHKSTW}.  We employ 
the lattice estimate for the ratio of color-singlet and color-octet components, 
$\mathcal{H}_8/\mathcal{H}_1$, listed in Eq.~(\ref{eq:H18}).  For the mixing angle, 
strong coupling strength, and $b$ quark mass, we use 
$\sin^2\theta=1/6$, $\alpha_s=0.2$, 
and $m_b=4.75$ GeV.  The sign of the product $\sin\theta\cos\theta$ is not 
constrained by experiment. We consider both possibilities; results for 
$\sin\theta\cos\theta > 0$ are shown in Fig.~\ref{figure:2}, and those for 
$\sin\theta\cos\theta < 0$ in Fig.~\ref{figure:3}.  The substantial differences 
in the $\chi_{b0}$ case between Fig.~\ref{figure:2} and Fig.~\ref{figure:3}
arise from the cancellation in the numerator of Eq.(\ref{eq:coefD0}) when 
$\sin\theta\cos\theta < 0$.  Owing to the phase space and angular momentum 
threshold suppression factors 
$|k|^{2J + 1}$, the SUSY contributions vanish as $m_{\tilde{b}} 
\rightarrow m_b$ ($|k| \rightarrow 0$) for all $\chi_{bJ}$ (as well as 
for the $\Upsilon$).  

The ratios of the decay rates through the bottom squark final state to the 
standard model prediction are shown in the second row of Figs.~\ref{figure:2} 
and ~\ref{figure:3}.  The dependence on the overall factor 
$\frac{4\pi\alpha_s^2 \mathcal{H}_1}{3m_b^4}$ 
cancels in this ratio, removing all dependence on the strong coupling strength 
$\alpha_s(m_b)$ (at the order in perturbation theory at which we 
work~\cite{comment}) and much of the dependence on the $b$ quark mass $m_b$ 
and on the matrix element $\mathcal{H}_1$. 
The contributions of bottom squarks decays to the hadronic widths are small 
in comparison to the standard model values except for the $\chi_{b0}$. 
As seen in Fig.~\ref{figure:2}, the SUSY contribution can be as large as 33\% of 
the standard model value for small $m_{\tilde{b}}$, $m_{\tilde{g}} = 12$ GeV, and 
positive $\sin\theta\cos\theta$.  The fraction decreases as $m_{\tilde{b}}$ and 
$m_{\tilde{g}}$ increase, and is reduced if $\sin\theta\cos\theta < 0$.  By 
contrast, for small $m_{\tilde{b}}$, the SUSY contributions are at most 15\% 
and 1\% of the standard model values in the $\chi_{b1}$ and $\chi_{b2}$ cases.  
Inspection of Eqs.~(\ref{eq:chi-SM}c) and (\ref{eq:GChi}) provides an easy 
explanation for the small role of the SUSY contribution in the $\chi_{b2}$ case.  
We observe that $\Gamma^{\rm SM} \sim (4/15) \Gamma_o$ whereas 
$\Gamma^{\rm SUSY} \sim D_8 m^2_b \Gamma_o \mathcal{H}_8/\mathcal{H}_1  
\sim (1/20) D_8 \Gamma_o$.  Since $D_8 \sim 1/16$, we obtain 
$\Gamma^{\rm SUSY}/\Gamma^{\rm SM} \sim$ 1\%.  

We comment on results obtained if the color-octet configuration is absent or, 
equivalently, the matrix element $\mathcal{H}_8$ is much smaller than estimated 
in Eq.~(\ref{eq:H18}).  The conclusions regarding $\chi_{b0}$ are essentially 
unchanged since the color-singlet contribution is dominant for both the SM and 
bottom squark decays.  For $\chi_{b1}$, the SM color-singlet term is absent at 
leading order; the SUSY decay mode would therefore dominate at this order.  
In the $\chi_{b2}$ case, the color-singlet rate for decay into bottom squarks 
is negligible compared to the SM color-singlet rate for $gg$ decay.  Overall, 
in the absence of color-octet contributions the predictions for the ratio 
$\Gamma_{\rm SUSY}/\Gamma_{\rm SM}$ would be qualitatively similar in the 
$\chi_{b0}$ and $\chi_{b2}$ cases to those in Figs.~\ref{figure:2} 
and~\ref{figure:3}, but very different for the $\chi_{b1}$.  

A ratio $R$~\cite{BBL-P} may be formed in which the color-octet contributions 
cancel:  
\begin{eqnarray}
R\equiv\frac{\Gamma(\chi_{b0})-\Gamma(\chi_{b1})}
            {\Gamma(\chi_{b2})-\Gamma(\chi_{b1})} ,
\label{eq:ratio}
\end{eqnarray}
This ratio depends only on the short-distance coefficients.  Results are shown in 
the third row of 
Figs.~\ref{figure:2} and ~\ref{figure:3}.  The ratio is enhanced significantly 
by the SUSY contribution for small $m_{\tilde{b}}$, $m_{\tilde{g}} = 12$ GeV, and 
positive $\sin\theta\cos\theta$.

Direct observation of $\chi_b$ decay into bottom squarks requires an understanding 
of the ways that bottom squarks may manifest themselves~\cite{BHKSTW,ELBrev}.  If 
the $\tilde{b}$ lives 
long enough, it will pick up a light quark and turn into a $J = 1/2$ $B$-mesino, 
$\widetilde{B}$, the superpartner of the $B$ meson.  The mass of the mesino could 
fall roughly in the range $3$ to $6$ GeV for the interval of $\widetilde{b}$ masses 
favored by the analysis of the bottom quark cross section. 
Charged $B$-mesino signatures in $\chi_{bJ}$ decay include 
single back-to-back equal momentum tracks in the center-of-mass; measurably lower 
momentum than lepton pairs; an angular distribution consistent with the decay 
of a state of spin $J$ into two fermions; and ionization, time-of-flight, 
and Cherenkov signatures typical of a particle whose mass is heavier than that 
of a proton.  On the other hand, possible baryon-number-violating R-parity-violating 
decays of the bottom squark lead to $u+s$, $c+d$, and $c+s$ final states.  These final 
states of four light quarks may be indistinguishable from conventional hadronic final 
states mediated by the two-gluon intermediate state.  Since the bottom squark 
decay mode is predicted to be substantial for $\chi_{b0}$'s, decays of the 
$\chi_{b0}$ primarily via the $\widetilde{B}$ possibility would offer an 
excellent opportunity to observe bottom squarks or to place significant limits 
on their possible masses.  

\begin{acknowledgments}
We acknowledge valuable conversations with Geoff Bodwin, Zack Sullivan, 
and Tim Tait.  Work in the High Energy Physics Division at Argonne National 
Laboratory is supported by the U.~S.~Department of Energy, Division of High Energy 
Physics, under Contract No.~W-31-109-ENG-38. 
\end{acknowledgments}

\begin{thebibliography}{}
\bibitem{comment}
Since the next-to-leading order results in Refs.~\cite{P-NLO} 
and~\cite{Petrelli} do not agree, we quote only leading order results 
in Eq.~(\ref{eq:chi-SM}).
\end{thebibliography}

\end{document}